# Two-Path Phonon-Interference Resonance Induces a Stop Band in Silicon Crystal Matrix by Embedded Nanoparticles Array

Shiqian Hu[1#], Lei Feng[1#], Shao Cheng[1], Yuriy A. Kosevich[2,3*], Junichiro Shiomi[1*]

[1] Department of Mechanical Engineering, The University of Tokyo, 7-3-1 Hongo, Bunkyo, Tokyo 113-8656, Japan

[2] N.N. Semenov Federal Research Center for Chemical Physics of Russian Academy of Sciences, 4 Kosygin Str., Moscow 119991, Russia

[3] Plekhanov Russian University of Economics, 36 Stremyanny per., Moscow 117997, Russia

[#] These authors contribute equally to this work.

[*] Corresponding author:

Email: yukosevich@gmail.com (Y.A.K.)

Email: shiomi@photon.t.u-tokyo.ac.jp (J. S.)

**Abstract**

In this work, we report a new stop-band formation mechanism by performing the atomistic Green's function calculation and the wave-packet molecular dynamics simulation for a system with germanium-nanoparticle array embedded in a crystalline silicon matrix. When only a single nanoparticle is embedded, the local resonance, induced through destructive interference between two different phonon wave paths, gives rise to several sharp and significant transmittance dips. On the other hand, when the number of embedded nanoparticles further increases to ten, a stop band with complete phonon reflection is formed due to the two-path resonance Bragg-like phonon interference. The wave packet simulations further uncover that the stop band originates from the collective phonon resonances at the embedded nanoparticles. Compared with the traditional stop-band formation mechanism that is the single-path Bragg reflection, the resonance mechanism has a significant advantage in not requiring the strict periodicity in the embedded nanoparticles array. We also demonstrate that the stop band can significantly suppress thermal conductance in the low-frequency regime. Our work provides a robust, scalable, and easily modulable stop-band formation mechanism, which opens a new degree of freedom for phononics-related heat control.

The ultimate impedance of phonon transport with a broad band of frequencies or mean free paths are of great scientific interest and engineering importance. This study is in response to the need for ultralow thermal conductivity crystalline materials in thermal management and thermal-energy storage/conversion technologies. Classical approaches to suppress thermal conductivity of crystalline materials have been aimed to introduce defects[1], impurities[2] and grain boundaries[3] to scatter phonons diffusively in the phonon-particle picture. For common crystalline materials, because the phonon scattering rate sensitively depends on the frequency as $\omega^{\alpha}$ ($\alpha>1$)[4], high-frequency phonons are more easily impeded compared to low-frequency phonons. In other words, as the frequency decreases, the transport of phonon becomes ballistic with equal unit energy transmission[5]. Therefore, manipulation of the lower frequency phonons is the key for further suppression of the thermal conductivity.

A promising approach to suppress the low-frequency phonon is to use the wave nature of phonons. There have been increasing number of reports on phononic crystals to harness interference[6-11] and/or resonance[12-17] of phonon waves to realize the phonon transport control beyond that based on the phonon-particle picture. Furthermore, since the phononic crystals require global periodicity, some of the recent studies have turned to local nanostructures to utilize local phonon resonance. So far reports on nano-pillars[12], nano-junctions[18], and nanoparticles[19] have unveiled unambiguous potential in reducing the thermal conductivity. These structures are practically superior to phononic crystals in terms of mitigating the harsh requirement to fabricate globally periodic nanostructure. Moreover, the local resonance allows us to manipulate the phonons with a wavelength much larger than the characteristic length of the nanostructure through interacting with the propagation phonons by creating a set of resonant frequencies[13, 14].

From the experimental perspective, the exciting progress has been made in embedding nanoparticles in silicon (Si) crystal matrix[20, 21]. Remarkably, the size of the nanoparticles can be as small as ~ 2 nm, which provides an ideal platform to modify and control the low-frequency phonon propagation and transport properties by the phonon local resonance effect. However, the local resonance originated from single nanoparticle only affects phonons with specific discrete frequencies, and the controllability of the overall thermal conductivity is limited. Therefore, there is a need for finding a local phonon-resonance phenomenon that can influence a broader and continuum range of frequencies, namely formation of a "stop band" in the phonon transmission function.

In this work, using the atomistic Green's function method, we explore the possibility to create the stop band by introducing the collective two-path phonon-interference resonance in a Si crystal matrix by embedding in it an array of germanium (Ge) nanoparticles. The spherical Ge nanoparticles force the phonons to propagate in two wave paths that are through and around the nanoparticles and destructively interfere with each other, yielding a stop band. The frequency range of the stop band can be modulated via varying the size of the nanoparticles. A significant advantage of the two-path phonon-interference resonances is that they no longer require strict periodicity, which is demonstrated here by introducing random displacements of the nanoparticles in the array. Finally, the impact of the resonance on the thermal conductance is investigated.

The atomistic Green's function (AGF) method is used to probe the ballistic phonon transport properties of the Si crystal matrix with several embedded Ge nanoparticles. Figure 1(a) shows a representative configuration of the AGF simulation: two semi-infinite leads sandwiching the scattering region. The scattering region is composed of several unit cells (dashed box in Fig. 1(a)), which are formed by embedding nanoparticles in the Si crystal. The unit cell is characterized by the side length $w$ and nanoparticle diameter $d$. When varying the nanoparticle diameter (varying from 1.1 to 1.6 nm) to investigate the influence of the diameter on the stop-band characteristics, we fix the area fraction ($\pi d^2 / 4w^2$). In the AGF method, the phonon transmission function can be calculated as[22-24]

$$T_p = Tr[\Gamma_L G^r \Gamma_R G^a] \qquad (1)$$

where $G^r$ and $G^a$ are the retarded and advanced Green's functions including the two leads, and $\Gamma_{\beta=L,R} = i(\sum_\beta^r - \sum_\beta^a)$ is the coupling function of the $\beta$ lead. Then, based on Landauer's theory, the phonon thermal conductance $G$ is given by

$$G(T) = \frac{h}{4\pi^2 S} \int_0^\infty \frac{\partial f_p(\omega)}{\partial T} T_p[\omega] \omega d\omega, \qquad (2)$$

where $f_p(\omega)$ is the Bose-Einstein distribution function for the phonons. For the force field, we adopt the Stillinger-Weber potential[25] of Si parameterizations describing the covalent bonds between Si-Si and Si-Ge atoms. Note, for the sake of simplicity, Ge is treated as a "heavy Si" with only the mass being the difference from Si, because the interatomic force constants of Si and Ge are transferable[26]. In addition, the lattice mismatch between Si and Ge has a negligible effect on the phonon transmission at Si/Ge interface[27].

We first study the possibility to form the stop band in the Si crystal matrix with the embedded Ge nanoparticles. Figure 1(b) shows the phonon transmission through the system at low-frequency part (<2.5 THz) for various numbers of Ge nanoparticles from zero (no particles) to ten. Here, the diameter of the nanoparticle is fixed at 1.1 nm, considering that the dominated thermal phonons wavelengths in pristine Si crystal are about 1~2 nm at room temperature[28].

The results show that, compared with the pristine Si crystal matrix (no particles), several local dips (1.9 and 2.1 THz) appear in the phonon transmission when the nanoparticles are embedded, due to the distinct local phonon resonance. This local phonon resonance phenomenon has been found in the previous theoretical[5, 10-12, 17] and experimental works[29]. Now, as the number of the nanoparticles increases, these discrete dips gradually merge and form a much wider and deeper dip, eventually resulting in a stop band when the number of the nanoparticle becomes as large as ten. This is similar to the Hong-Ou-Mandel dip in quantum optics[30-33] in that the phonons propagating through two different paths interfere destructively (inset in Figure 1(d)). In addition, another local dip appears at around 1.1 THz (TA mode) as the number of the nanoparticles increases, which originates from the collective phonon resonances of the nanoparticles (details are shown in supporting information Fig. S1). In our simulations, we set the nominal distance between the adjacent nanoparticles ($L_d$) as 2.2 nm, which is half the wavelength $\lambda_R$ in Si crystal matrix of the phonon corresponding to the resonance dip at a single nanoparticle (at 1.9 THz), to generate the two-path resonance Bragg-like multiple reflection from local phonon scatterers[10]. We also simulated another two systems with different $L_d$ (1.65 nm and 2.75 nm) which do not satisfy the two-path resonance Bragg-like condition, leading to the disappearance of the stop band (details are shown in supporting information Fig. S2). Note that although this may sound as if we need to position the nanoparticles with specific periodicity, that is not the case. The positions of nanoparticles themselves can be non-periodic as long as the unit cells satisfy Bragg's condition, as will be discussed later.

To understand the underlying physical mechanism for the formation of the stop band, we also performed the phonon wave-packet simulations, which can directly capture the wave characteristics of phonon propagation in the coordinate space[34-36] (calculation details are shown in supporting information Part I). For comparison, we perform the wave-packet simulations with (1 and 10 nanoparticles) and without nanoparticles. For each structure, a wave packet with a specific mode and length is formed and launched in the pristine Si part, and its propagation process is further recorded along the length direction of the system. Here, the longitudinal acoustic phonon with the frequency of 1.9 THz is launched, which corresponds to the first transmission dip in the Si crystal matrix with one nanoparticle (Fig. 1(b)). The coherence length of the wave packet is set as 20 nm, much longer than the wavelength of the central phonon.

Figure 2 shows the wave-packet propagation process in the pristine Si crystal matrix (no particles) and Si crystal matrix with one and ten embedded Ge nanoparticles. It is noteworthy that the wave packets are modeled with a very small amplitude ($2\times10^{-4}$ nm) in order to avoid involving the lattice anharmonicity. Therefore, as we expected, the wave packet travels ballistically in the pristine Si crystal matrix due to the negligible anharmonic phonon scattering (Fig. 2(a)).

As shown in Fig. 2(b), in the Si crystal matrix with one Ge nanoparticle, the original wave packet separates into two waves (reflected and transmitted waves) after impacting on the nanoparticle. These two waves further propagate independently in opposite directions, which indicates incomplete phonon transmission. What is noteworthy here is that the nanoparticle keeps vibrating even after the wave packet has passed away (red dashed box in Fig. 2(b)), which confirms the presence of the local resonance in the real space. In addition, the time evolution of the center of mass (COM) of nanoparticle presents temporal vibration with a period corresponding to the resonant frequency (1.9 THz), which confirms that the nanoparticle resonates with the incident phonons (details in the supporting information Part II).

In the case of ten nanoparticles, the wave packet is nearly totally reflected (stop-band formation) by the nanoparticles array. The evolution of the wave packet in Fig. 2(c) shows that the nanoparticles resonance is significantly enhanced with a larger amplitude of local vibration (red dashed box in Fig. 2(c)). During the vibration, the adjacent nanoparticles are displaced in opposite directions along the *x*-axis (inset in Fig. 2(c)), clearly demonstrating the out-of-phase collective phonon resonance for nearest nanoparticles (details in the supporting information Part II). Therefore, the increase in the number of nanoparticles gives rise to the collective phonon resonance in the two-path phonon interference, which leads to the stop-band formation.

The previous report showed that, in the case of a single nanoparticle, the resonance effect is significantly deteriorated unless the particle size is two-orders-of-magnitude smaller than the coherence length[19], which limits the controllability of thermal transport in Si crystal. Our current finding (collective phonon resonance) suggests that the transmission dip can be further enhanced, up to the formation of the stop band, by simply increasing the number of the nanoparticles with the proper average inter-particle spacing. This enhancement opens up an elevated opportunity for engineering thermal transport, especially considering that the inhibition of low-frequency phonons has been the key challenge[37].

Since the local resonance frequencies are closely related to the diameter of the nanoparticles[19], we further explore the diameter effect on the stop-band formation. The result shows that the stop band presents a distinct red-shift as the diameter of the nanoparticles increases (Fig. 3(a)), which is consistent with the shift observed in the previous works[37]. Generally, the phonon scattering rate depends on frequency as $\sim \omega^{\alpha}$ ($\alpha$>1)[4, 38], which means that the high-frequency phonons can be easily blocked while the low-frequency phonons are hardly altered. The flexibility of tuning the stop band in low-frequency regime by manipulating the two-path phonon interference provides a new avenue for engineering the thermal properties of materials. Furthermore, very recently, there has been great progress in thermal rectification based on the graded structure[39-42]. In a similar mind, a more broadening of stop band could be expected in a graded Si matrix where the diameters of nanoparticles are continuously increased if the multi-size two-path phonon interference works.

The above demonstration of the stop-band formation and frequency tuning was done for an ideal case where the nanoparticles are placed with an equidistance. However, fabricating such an embedded structure with a periodic array of nanometer-size (1-2 nm) particles would be difficult even with the state-of-art nanotechnology. In fact, one of the largest merits of the current two-path phonon interference is that a strict periodicity is not required, unlike the traditional single-path phonon Bragg interference used in superlattices. In superlattices, as illustrated in Fig. 1(c), the multiple reflected waves induced by the sharp interfaces between the alternating different materials interfere constructively and thus prevent the incident wave from propagating within the structure. On the other hand, the current collective resonance illustrated in Fig. 1(d) does not require reflection at sharp interfaces, and thus should be less sensitive to the positioning of the local resonance scatterers. To confirm this aspect, we simulate the stop-band formation by introducing the random displacements to the position of the nanoparticles in each unit cell. The inset in Fig. 3(b) shows the cross-section figure of the unit cell for the Si crystal matrix. The red dot denotes the center of the nanoparticle. The six orange dots on the black cycles denote the possible center of the nanoparticles when we explore the random-displacement effect on the stop-band formation.

Here, the positions of the nanoparticles are displaced by a distance of $L_s$ from the original regular periodic positions in random directions. Figure 3(c) shows a schematic figure of the simulation domain when the random displacements are introduced. We performed three independent simulations for different $L_s$ (0.1, 0.3 and 0.5 nm) to show how the magnitude of displacement affects the stop-band formation. The result (Fig. 3(b)) shows that the stop band is well maintained regardless of the random displacement: only the average inter-particle spacing fulfilling the two-path resonance Bragg-like condition is important. This clearly indicates that the stop band formed by the two-path phonon interference is extremely robust and easy to implement, compared with the traditional mechanism (single-path Bragg multiple reflection) that requires the strict periodicity.

We further explore the stop-band effect on the thermal conductance of the Si crystal matrix. Figure 4 (a) and 4 (b) shows the spectral thermal conductance of pristine Si crystal (no particles) and Si crystal matrix with embedding one nanoparticle and ten nanoparticles at gamma point and full Brillouin zone at 300 K, respectively. Here, the diameter of the nanoparticles is fixed at 1.1 nm. As shown in Fig. 4 (a), we find that the thermal conductance of pristine Si crystal is significantly suppressed when the nanoparticles are introduced. This frequency-dependent suppression corresponds well with the local resonance frequency in phonon transmission. Besides, as the number of the nanoparticles increases, the thermal conductance is further reduced especially in the stop-band region. In contrast, the resonance features are concealed in that of the full Brillouin zone case because many other phonon modes with different wave vectors are superimposed (Fig. 4(b)). However, we note that comparing with the pristine Si crystal, the thermal conductance of the phonons with a frequency above 1 THz show a significant reduction, especially for the ten nanoparticles case, which reflects the resonance effect on the thermal transport. Moreover, the normalized thermal conductance for the full Brillouin zone at low-frequency range (< 3THz) with respect to the pristine Si crystal shows that a single nanoparticle can give a reduction ratio of 16.3%, and this is significantly enhanced in the ten particles case to 41.8%.

Finally, we briefly introduce the research significance and its potential application. The stop-band formation originating from the distinct local resonance in two interfering phonon paths breaks down the traditional restriction that requires a strict periodicity of the structure. This new mechanism suggests that a robust stop band can be achieved by embedding nanoparticles in the Si nanocrystal matrix, which is much easier to fabricate and scale-up in practical experiments. Furthermore, the effective modulation of the stop band can be combined with other thermal transport manipulation mechanism to scatter phonons by impurities, disorders, and grain boundaries to cover the entire heat conduction spectrum to realize ultimate inhibition of heat conduction. This also speeds up the development of the high-performance phononic devices by controlling thermal energy flow by offering a more practical path to utilize phonon coherence. This may lead to the phononic devices for controlling heat, such as heat waveguide, thermal imaging, thermal rectifier and thermal storage, in analogy with the photonics for electromagnetic waves and phononics for sound. There, the recent advanced possibilities to optimize the structures for phonon transport applications using materials informatics[43, 44] (machine learning) will be useful.

In summary, we have reported a new class of stop-band formation mechanism that is two paths phonon interference based on Si crystal matrix with embedded nanoparticles array by performing the atomistic Green's function calculations. We showed that the stop-band position is closely related to the diameter of the nanoparticles while the randomness has nearly no effect on the stop-band formation. By using the wave-packet simulation analysis, we demonstrated that the stop band arises from the two-path phonon interference induced by distinct local resonance, which is completely different from the traditional superlattice structure. Finally, a significantly suppressed thermal conductance in the low-frequency part induced by the stop band is presented. The advantage of this stop-band formation mechanism is that it no longer requires strict periodicity, which makes the effective-modulation largely accessible, and thus the work offers the possibility for a further significant development of the phononic-related devices.

**Acknowledgments**

This research was funded in parts by JSPS KAKENHI Grant Nos. 18F18058 and 19H00744, by RFBR and JSPS according to the research project No. 19-52-50030, and by JST CREST Grant No. JPMJCR16Q5. S. H. and J. S. acknowledge support from Japan Society for the Promotion of Science (JSPS) Postdoctoral Fellowship for Overseas Researchers Program (P18058). L. F. acknowledges support from JSPS DC Fellowship. Y.A.K. acknowledges support from the Russian Science Foundation through award No. 16-13-10302 (for the analysis of the two-path resonance Bragg condition) and from Invitation Fellowships for Research in Japan by JSPS and CREST.

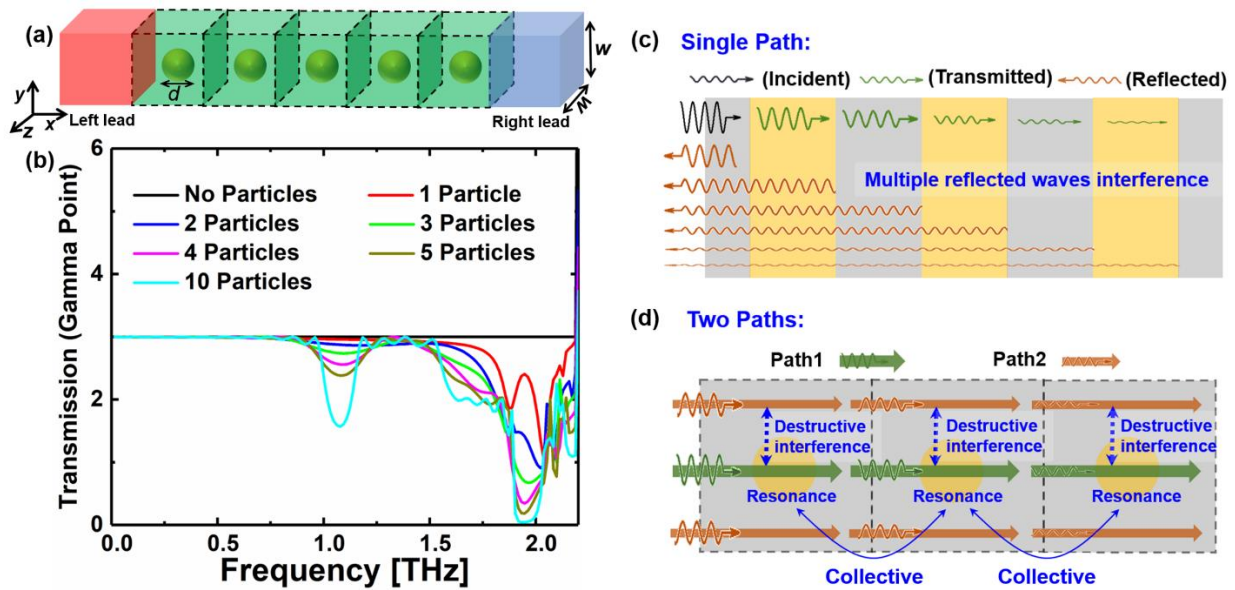

**Figure 1. Schematic simulation set up of Si crystal matrix with embedded Ge nanoparticles array and the corresponding phonon transmission spectrum with different numbers of nanoparticles at low-frequency range and the schematic figures for the traditional Bragg phonon interference (single path) and the two-path phonon interference.** (a) A schematic picture of Si crystal matrix with embedded nanoparticles array. (b) Low-frequency part of the phonon transmission spectrum for the Si crystal matrix with embedded 0-10 nanoparticles. (c) An illustration of the traditional Bragg phonon interference in the superlattice (multiple reflected waves interfere constructively and thus the incident wave is fully reflected). (d) An illustration of the destructive interference between two phonon paths (through and around the nanoparticles).

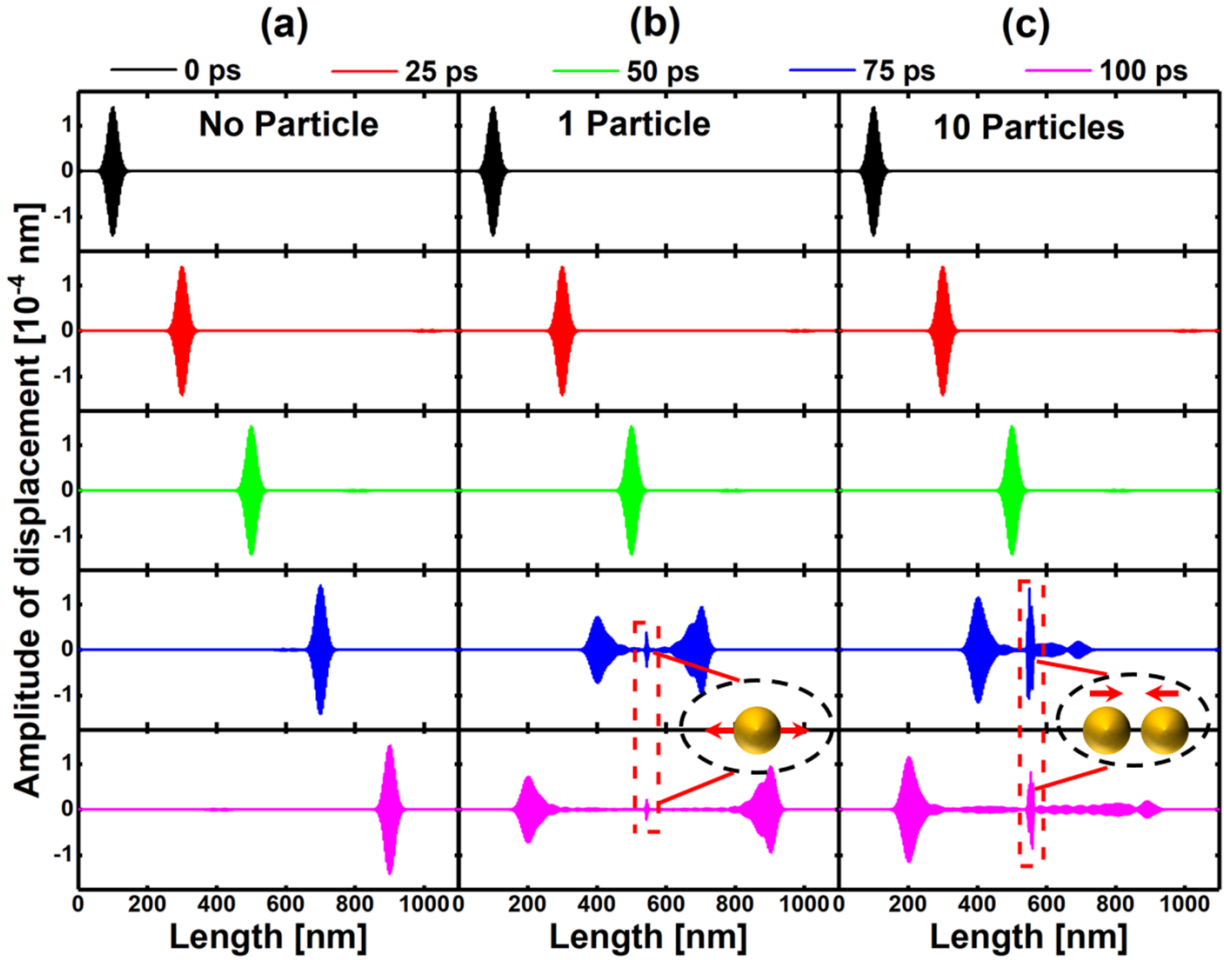

**Figure 2. The wave packet simulations in different Si crystal matrix.** Snapshots of displacement for an LA phonon wave packet in (a) pristine Si crystal matrix (no particle), (b) Si crystal matrix with one Ge nanoparticle, and (c) Si crystal matrix with ten nanoparticles. The red dashed boxes in (b) and (c) denote the position where the local resonance occurs in the Si crystal matrix with the embedded nanoparticles. The inset figures in (b) and (c) illustrate the motions of the nanoparticles. The red arrows denote the vibrational direction.

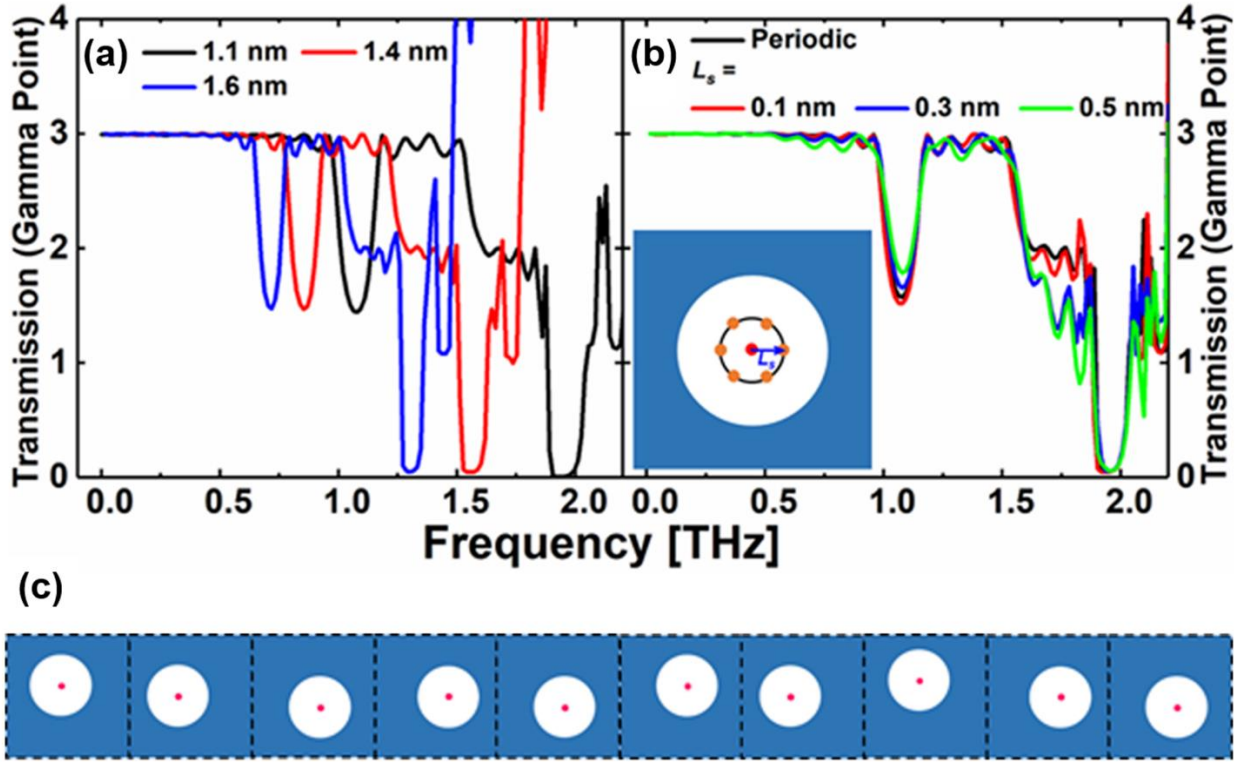

**Figure 3. The effects of diameter and random displacement of the nanoparticle on the stop-band formation.** (a) The dependence of stop-band formation on the diameter of the nanoparticles. Here, the number of nanoparticles is fixed at 10. The diameter of the nanoparticles ranges from 1.1 nm to 1.6 nm. (b) The effect of random displacements of nanoparticles away from the regular periodic positions on the stop-band formation. The inset shows the cross-section of the unit cell. The red dot denotes the center of the nanoparticle. The six orange dots denote the chosen possible centers of the nanoparticles when the nanoparticles are displaced in random directions by a distance of $L_s$. (c) A schematic picture of the simulation domain when the random displacements of the nanoparticles from their regular periodic positions are introduced.

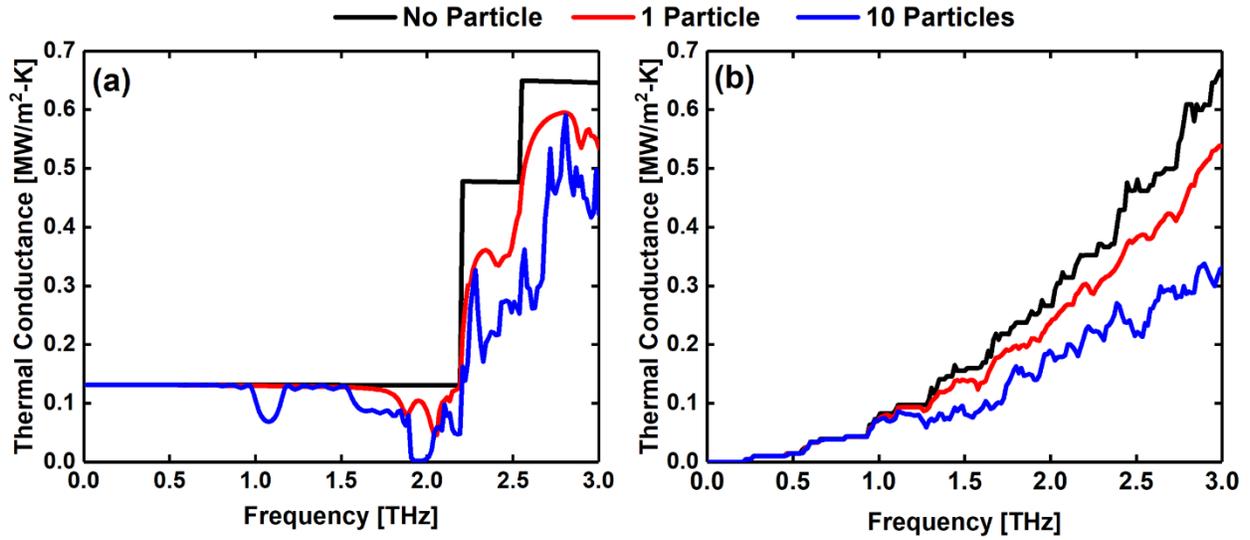

**Figure 4. Spectral thermal conductance at low-frequency range.** Spectral thermal conductance of pristine Si crystal without nanoparticles (black), Si crystal matrix with one Ge nanoparticle (red) and ten Ge nanoparticles (blue) considering (a) only wave-vectors with $k_y=k_z=0$ and (b) all the wave-vectors in the first Brillouin zone, respectively. The temperature is 300K.

# Supporting information

**Two-Path Phonon-Interference Resonance Induces a Stop Band in Silicon Crystal Matrix by Embedded Nanoparticles Array**

Shiqian Hu[1#], Lei Feng[1#], Shao Cheng[1], Yuriy A. Kosevich[2,3*], Junichiro Shiomi[1*]

[1]Department of Mechanical Engineering, The University of Tokyo, 7-3-1 Hongo, Bunkyo, Tokyo 113-8656, Japan

[2] N.N. Semenov Federal Research Center for Chemical Physics of Russian Academy of Sciences, 4 Kosygin Str., Moscow 119991, Russia

[3] Plekhanov Russian University of Economics, 36 Stremyanny per., Moscow 117997, Russia

[#]These authors contribute equally to this work.

[*]Corresponding author:

Email: yukosevich@gmail.com (Y.A.K.)

Email: shiomi@photon.t.u-tokyo.ac.jp (J. S.)

## I. Phonon wave packet simulation method

For the phonon wave packet simulation [1-3], the diameter and side length are set as 1.1 and 2.2 nm respectively. Periodic boundary conditions are used in all directions. The wave packet was formed via linear combinations of the vibration normal modes as following

$$\mu_{il,\alpha} = \frac{A}{m_i} \varepsilon_{i\alpha,\lambda} e^{ik_0(x_l - x_0)} e^{-(x_l - x_0)^2/\eta^2}, \tag{S1}$$

where $\mu_{il,\alpha}$ is the $\alpha$th displacement component of $i$th atom in the $l$th unit cell, $A$ is the amplitude of the wave packet, $m_i$ is the mass of the $i$th atom and $\varepsilon_{i\alpha,\lambda}$ is the $\alpha$th eigenvector component of eigenmode $\lambda$ for the $i$th atom. The wave packet has the wave-vector $k_0$ and is centered around $x_0$ in the coordinate space. The parameter $\eta$ is used to define the spatial width (coherence length) of the wave packet. To initialize velocities, we add time dependence to Eq. S1 and differentiate it as,

$$v_{il,\alpha} = -i\omega_\lambda u_{il,\alpha}, \tag{S2}$$

To present the phonon collective local resonance behavior in the Si crystal matrix, we perform the wave packet simulation and record the coordinate displacement of Ge nanoparticles in the time domain (Fig. S3). Here, the amplitude $A$ is fixed at 0.002Å.

## II. Wave packet simulation results and the analysis of the vibrational mode

For comparison or contrast, we firstly present the propagation process of wave packet of an LA phonon mode (0.45 THz) in the Si crystal matrix with ten embedded nanoparticles. As shown in Fig. S3(a), the wave packet propagates ballistically in the simulation system, presenting a complete transmission behavior (consistent with the AGF result shown in Fig. 1(b)). The corresponding time evolution of the center of mass (COM) of nanoparticles first increases and then decreases as the wave packet passes through the nanoparticles (Fig. S3(b)).

However, as shown in Fig. S3(c), for the Si crystal matrix with one embedded nanoparticle, when the phonon mode of the wave packet is set as 1.9 THz, the nanoparticle remains vibrating even after the wave packet has passed away and the vibrational amplitude is remarkably enhanced (in comparison with the amplitude of the incident wave packet, $2\times10^{-4}$ nm). In addition, the time evolution of the COM of nanoparticle presents a temporal vibration period corresponding to the resonant frequency (1.9 THz). These results together confirm that the nanoparticle resonance with the incident phonons occurs.

When the number of Si nanoparticles embedded in the Si crystal matrix increases to ten, the nanoparticle resonance effect is enhanced further (Fig. S3(d)). Besides, a more striking finding is that the adjacent nanoparticles present an oppositely rattling motion along the x-axis (inset in Fig. S3(d)), which demonstrates a clear out-of-phase collective phonon resonance.

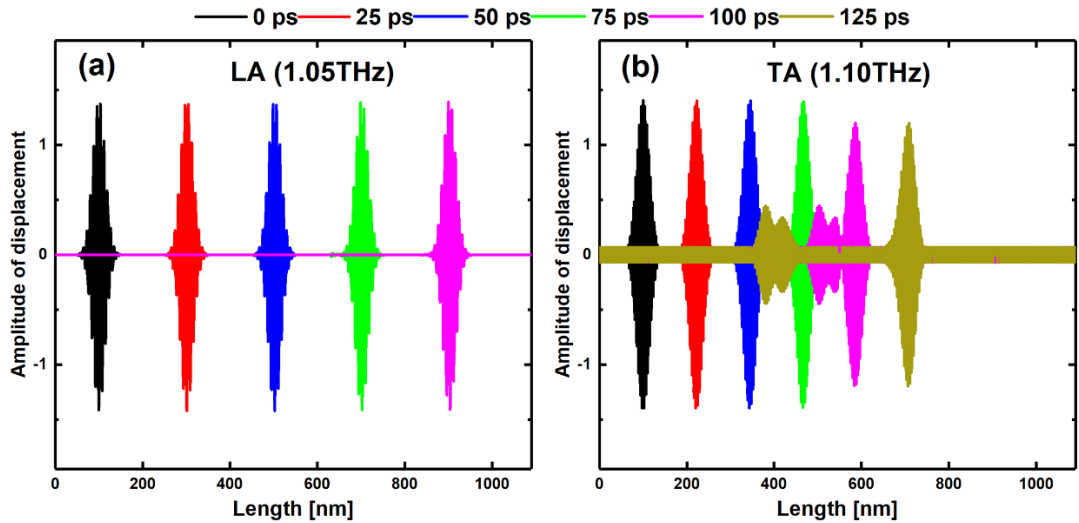

**Figure S1. The wave packet simulations of LA phonon (1.05 THz) and TA phonon (1.10 THz) in Si crystal matrix with ten embedded Ge nanoparticles** (a), (b) are the displacement snapshot of wave packets of LA phonons (with 1.05 THz central frequency) and TA phonons (with 1.10 THz central frequency) in the Si crystal matrix with ten embedded Ge nanoparticles, respectively. For the LA phonons, the wave packet travels ballistically in the Si crystal matrix. In contrast, for the TA phonons, the wave packet separates into the reflected and transmitted waves after impacting on the nanoparticle. These two figures together confirm that the dip at around 1.1 THz in Fig. 1(b) corresponds to the TA phonon mode.

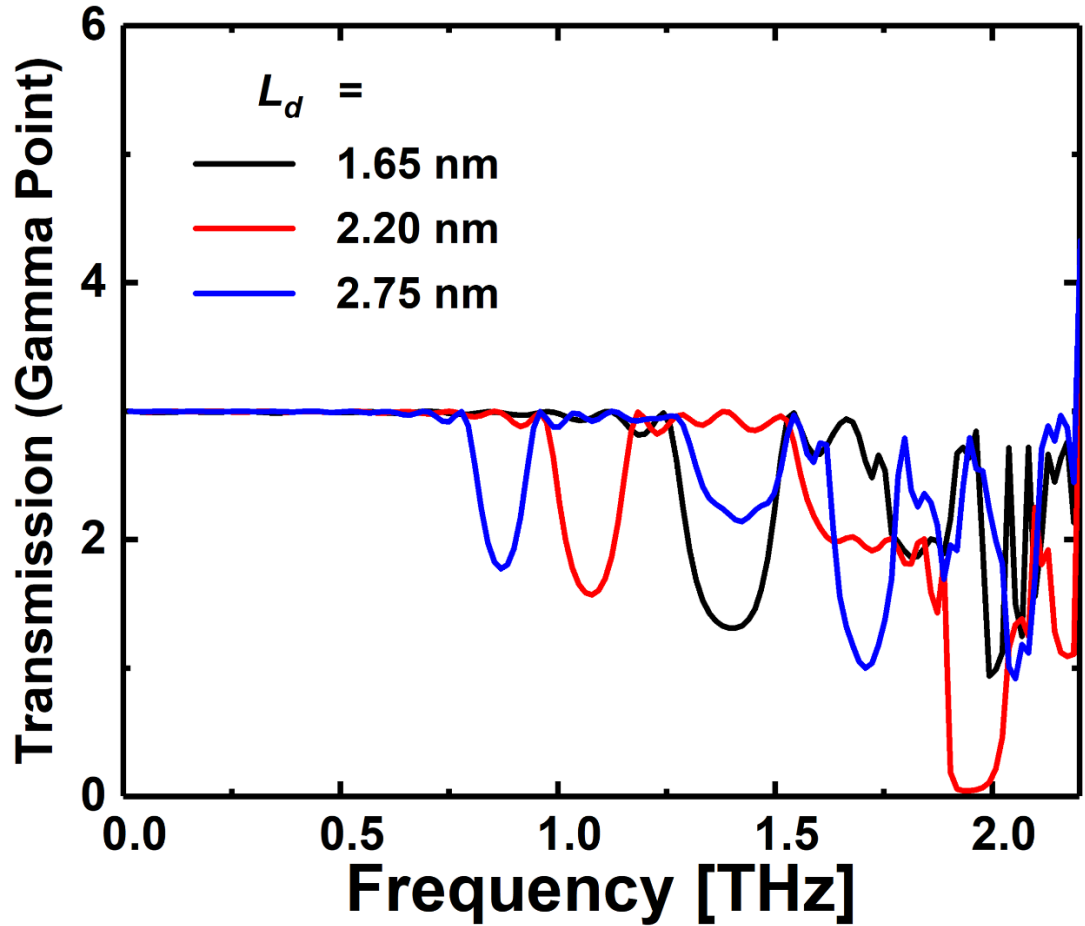

**Figure S2. The phonon transmission spectrum in the Si crystal matrix with ten embedded Ge nanoparticles with different distances between the adjacent nanoparticles ($L_d$) in low-frequency range.** Low-frequency part of the phonon transmission for the Si crystal matrix with ten embedded nanoparticles with different $L_d$. The stop band disappears when the $L_d$ does not satisfy the two-path resonance Bragg-like condition ($L_d = 1/2\ \lambda_R$).

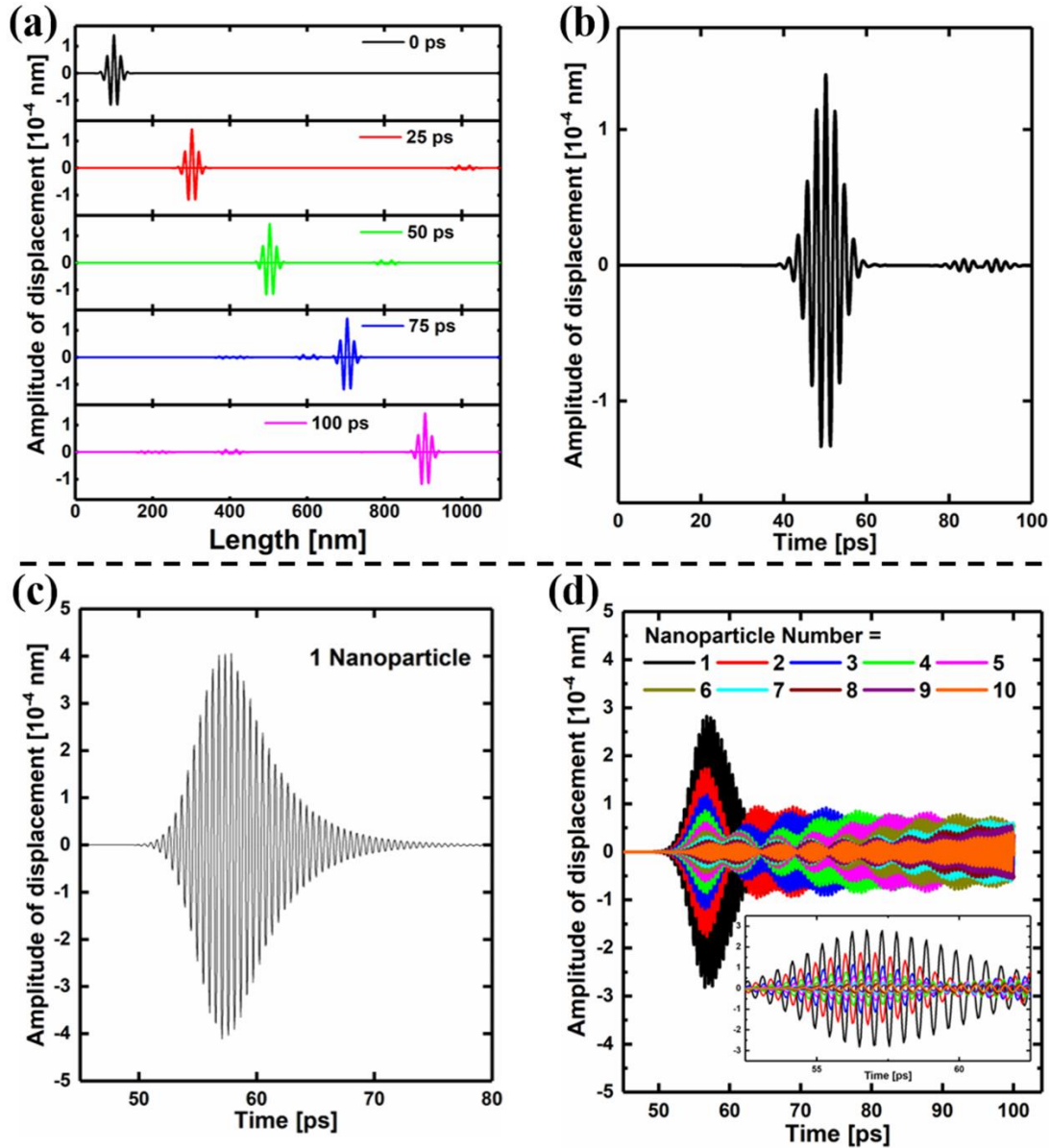

**Figure S3. The wave packet simulations for LA phonons in Si crystal matrix with ten embedded Ge nanoparticles and the time-evolution of the center of mass (COM) of nanoparticles with different frequencies.** (a) Displacement snapshots of wave packet for LA phonons (with 0.45 THz central frequency) in Si crystal matrix with ten embedded nanoparticles. (b) The corresponding time evolution of x-coordinate displacement of the COM of the nanoparticles with the frequency 0.45 THz. (c), (d) are the time evolution of x-coordinate displacement of the COM of the nanoparticles with the frequency 1.9 THz in the Si crystal matrix

with one and ten embedded nanoparticles, respectively. The inset figures in (d) shows the displacement of the COM between 45 and 65 ps.